\documentclass[a4paper]{article}

\usepackage{INTERSPEECH2018}
\usepackage[none]{hyphenat}
\usepackage{multirow}
\usepackage{float}
\usepackage{amsmath}
\usepackage[linesnumbered,ruled]{algorithm2e}
\usepackage{stackrel}
\usepackage{sectsty}
\usepackage{comment}
\usepackage{subfloat}
\usepackage{cite}
\usepackage{hyperref}
\title{Semi-supervised Sequence-to-sequence ASR using Unpaired Speech and Text}
\name{Murali Karthick Baskar$^{\phi,\dagger}$, Shinji Watanabe$^\dagger$, Ramon Astudillo$^\pi$, Takaaki Hori$^\Upsilon$, \newline Luk\'{a}\v{s} Burget$^\phi$, Jan \v{C}ernock\'{y}$^\phi$
}
\address{
  $^\phi${Brno University of Technology, Czech Republic},\\
  $^\pi${IBM T.~J.~Watson Research Center, USA},\\
  $^\dagger${Johns Hopkins University, USA},\\
  $^\Upsilon${MERL, USA}}
  
\email{baskar@fit.vutbr.cz}

\begin{document}

\maketitle
\begin{abstract}
Sequence-to-sequence automatic speech recognition (ASR) models require large quantities of data to attain high performance. For this reason, there has been a recent surge in interest for unsupervised and semi-supervised training in such models. This work builds upon recent results showing notable improvements in semi-supervised training using cycle-consistency and related techniques. Such techniques derive training procedures and losses able to leverage unpaired speech and/or text data by combining ASR with Text-to-Speech (TTS) models. In particular, this work proposes a new semi-supervised loss combining an end-to-end differentiable ASR$\rightarrow$TTS loss with TTS$\rightarrow$ASR loss. The method is able to leverage both unpaired speech and text data to outperform recently proposed related techniques in terms of \%WER. We provide extensive results analyzing the impact of data quantity and speech and text modalities and show consistent gains across WSJ and Librispeech corpora. Our code is provided in ESPnet to reproduce the experiments.
\end{abstract}
\medskip
	
\noindent\textbf{Index Terms}:  Sequence-to-sequence, end-to-end, ASR, TTS, semi-supervised, unsupervised, cycle consistency

\section{Introduction}
Sequence-to-sequence (seq2seq) ASR training directly maps a speech input to an output character sequence using a neural network~\cite{graves2014towards,bahdanau2016end,chan2016LAS}, similar to those used in machine translation~\cite{bahdanau2014neural,sutskever2014sequence}. The model requires a considerable amount of paired speech and text to learn alignment and classification~\cite{amodei2016deep,prabhavalkar2017comparison}, which limits its use in under-resourced domains. On the other hand, unpaired speech and text can be obtained for most domains in large quantities making training with unpaired data particularly relevant for seq2seq models.

Recent works have shown that the problem of seq2seq training in low-resource conditions can be tackled using unpaired data. These methods can be classified into three categories according to the type of data used. First, methods utilizing only unpaired speech for unsupervised training. In this category,~\cite{tjandra2018end} proposes an end-to-end differentiable loss integrating ASR and TTS models by the straight-through estimator. The work in~\cite{hori2018cycle} also proposes an end-to-end differentiable loss integrating ASR and a Text-to-Encoder (TTE) model. Both works, showed that connecting ASR with TTS/TTE can handle unpaired speech data as well as reduce ASR recognition errors.
The second category concerns methods leveraging unpaired text data, which focus on enhancing the decoder component of seq2seq ASR~\cite{hayashi2018back} or moving the encoder representation closer to text representation~\cite{renduchintala2018multi}. The former approach~\cite{hayashi2018back} is implemented by feeding the text data to TTE-to-ASR~\cite{hayashi2018back} model, where the TTE converts the text directly to encoder representations and then is fed to the decoder. In the latter model, 
an encoder component is shared to have common representation across both speech and text data~\cite{renduchintala2018multi}. In addition to these works, \cite{liuLM} used a language model (LM) built with unpaired text-only data to jointly train with an ASR. In the third category, both unpaired speech and text data are exploited. Examples of this are the speech-chain framework~\cite{tjandra2017listening} and adversarial training schemes~\cite{drexler2018combining,karita2018sequence}. The former proposes a dual pipeline of ASR$\rightarrow$TTS and TTS$\rightarrow$ASR systems. Here, the backpropagation only takes place in the secondary stage of the chain. The latter utilizes, adversarial learning objectives to match the speech and text distributions.

Outside of the speech processing area, cycle-consistency can be related to successful progress in machine translation~\cite{he2016dual}, which derive end-to-end differentiability by using N-best approach. Also, improvements are observed in image processing tasks using cycle-consistency with an adversarial objective~\cite{zhu2017unpaired}.

In this paper, we borrow ideas from the above mentioned techniques to handle unpaired data using cycle-consistency. The basic idea of this approach is that if a model converts an input data to output data and another model reconstructs the input data from the output data, then the input data and its reconstruction should be similar. We use this idea to derive a new loss able to use both unpaired speech and text data. The contributions of this paper are the following
\begin{itemize}
    \item A speech-only end-to-end differentiable loss is proposed using ASR$\rightarrow$TTS. This improves the TTE cycle loss in \cite{hori2018cycle} and is related to the straight-through loss in~\cite{tjandra2018end}
    \item We complement this with a text-only pipeline loss using TTS$\rightarrow$ASR. This improves TTE backtranslation \cite{hayashi2018back} and is related to the speech-chain in \cite{tjandra2017listening}
    \item We combine this and a shallow integrated RNNLM~\cite{hori2017advances,kannan2018analysis} to obtain a high performance on unpaired data.
\end{itemize}
Our experimental analysis using WSJ and Librispeech corpus described in section~\ref{sec:4} substantiate our hypothesis that unpaired speech and text can individually improve ASR performance and combining them provides additional gains.

\section{Cycle consistency training}\label{sec:2}

\subsection{ASR and TTS with seq2seq models}
Sequence-to-sequence ASR systems \cite{bahdanau2016end} model directly a probability distribution over $C$-length character or phoneme sequence $\mathbf{Y}=\{y_c\}_{c=1}^C$ given $T$-length speech features $\mathbf{X}=\{x_t\}_{t=1}^T$ i.e. Mel-filterbank as 

\begin{equation}
p_{\mathrm{ASR}}(\mathbf{Y}\mid\mathbf{X})=\prod_{c}^{C}p(y_{c}\mid{y}_{1:c-1},\mathbf{X})\label{eq:pyX}.
\end{equation}

Each character prediction depends on the entire input and all previously generated characters. To account for this complex relation, a neural network with attention mechanism is used \cite{bahdanau2014neural}. This is composed of an encoder network, typically one or more bi-directional LSTMs layers \cite{hochreiter1997long}, generating an encoder sequence $\mathbf{H}=\{h_t\}_{t=1}^T$ as
\begin{equation}
\mathbf{H} = \mathrm{encoder}(\mathbf{X}).
\label{encoder}
\end{equation}
A decoder, also modeled by one or more LSTM networks, utilizes a matching function to select the relevant encoded input features$\{h_t\}_{t=t_c}^{t'_c}$ for each step $c$ given its internal state. This matching function is normalized to resemble a probability distribution over the input features and is termed attention. Decoder internal state, attended input and previously produced character are used to predict the current character. 

When paired data is available, the seq2seq models are trained with the cross-entropy (CE) criterion given correct output $y^{*}$:
%
\begin{align}
\mathcal{L}_{\mathrm{ASR}}&=-\mbox{log}\,p_{\mathrm{ASR}}(y^{*}\mid{X}) \\\nonumber &=-\sum_{l=1}\log p_{\mathrm{ASR}}(y_{l}^{*}\mid{y}_{1:l-1}^{*},X).\label{eq:3}
\end{align}

%
TTS seq2seq models use a very similar architecture, but receive the characters as input $\mathbf{Y}=\{y_c\}_{c=1}^C$ and predict the speech features with a regression layer $\mathbf{X}=\{x_t\}_{t=1}^T$. There are some needed modifications with respect to ASR, such as the need to predict continuous data and the end of utterance. In this work, we used the Tacotron2 architecture, detailed in \cite{shen2018natural}. The loss of the TTS system is divided into three terms:
\begin{equation}
\mathcal{L}_{\mathrm{TTS}}=\mathcal{L}_{\mathrm{MSE}} + \mathcal{L}_{L1} + \mathcal{L}_{BCE} \label{eq:4}
\end{equation}
where the mean square error (squared L2) and L1 norms are over a regression estimate $\hat{\mathbf{X}}$ and BCE is the binary cross entropy loss for the end of sentence prediction. MSE and L1 components of the loss can be interpreted as the minus log-probability of the speech features for Gaussian and Laplace distributions for constant scale parameters i.e.
\begin{align}
\mathcal{L}_{\mathrm{TTS}} &= -\log p_{\mathrm{TTS}}(\mathbf{X}^*\mid\mathbf{Y})\\\nonumber &= -\sum_{t=1}\log p_{\mathrm{TTS}}(\mathbf{X}^*_t\mid\mathbf{X}^*_{1:t-1}, \mathbf{Y})\label{eq:5}
\end{align}
considering the MSE only, we have
\begin{equation}
-\log p_{\mathrm{TTS}}(\mathbf{X}^*_t\mid\mathbf{X}^*_{1:t-1}, \mathbf{Y}) \propto\\ \|\mathbf{X}^*_t - \mathbf{g}(\mathbf{\mathbf{Y}}, \mathbf{X}^*_{1:t-1}) \|^2
\end{equation}
where $\hat{\mathbf{X}}_t = \mathbf{g}(\mathbf{Y}, \mathbf{X}^*_{1:t-1})$ is the Tacotron2 auto-regressive estimate at time $t$.

\subsection{Unsupervised training with unpaired speech data}

The previous formulations concern the case in which both paired speech $\mathbf{X}$ and text $\mathbf{Y}$ are available. 
In this section we propose an improvement over recent work in \cite{hori2018cycle} that provides an unsupervised loss for ASR while being more performant. A central problem to the ASR$\rightarrow$TTS pipeline is the fact that the text bottleneck eliminates a lot of information from speech e.g. speaker identity. The work in~\cite{hori2018cycle} solves this by training the TTS model to predict encoded speech, $\mathbf{H}$ in eq.~\eqref{encoder}, instead of speech $\mathbf{X}$ directly. Since the encoded speech keeps some speech characteristics, it is possible to define the following loss based on the Cycle-Consistency (CC) criterion

\begin{figure}[H]
\begin{centering}
\includegraphics[scale=0.5]{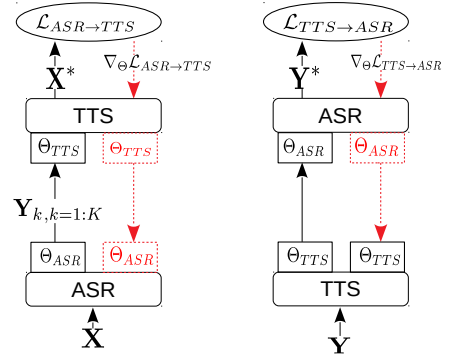}
\caption{Simplified representation of the cycle-consistency training for unpaired speech and text. The flow of the diagram is bottom-to-top. The red dotted lines denotes the gradient propagation and the updated parameters. Figure~\ref{fig:overview}a denotes the ASR$\rightarrow$TTS pipeline where $\mathbf{Y}_{k}$ is the $k^{\mathrm{th}}$ randomly sampled ASR output. Figure~\ref{fig:overview}b shows the TTS$\rightarrow$ASR pipeline.}\label{fig:overview}
\end{centering}
\end{figure}

\begin{equation}
\mathcal{L}_{\mathrm{ASR}\rightarrow\mathrm{TTE}}= E_{p_{\mathrm{ASR}}(\mathbf{Y}\mid \mathbf{X})}\{\mathcal{L}_{\mathrm{TTE}}\} \label{eq:6}
\end{equation}
where $\mathcal{L}_{\mathrm{TTE}}$ is the same as $\mathcal{L}_{\mathrm{TTS}}$ but replacing $\mathbf{X}$ by the encoded speech $\mathbf{H}$. This loss penalizes the ASR system for transcriptions that, once transformed back into an estimate of the encoded speech $\hat{\mathbf{H}}$ by the TTE, differ from the original encoded speech $\mathbf{H}$. Such loss does not require to have the correct text output $y^*$ and is end-to-end differentiable.


TTE-CC has the disadvantage of having to train an specific network to predict $\mathbf{H}$. The encoded speech, may also already eliminate some of the speech characteristics making the CC loss less powerful. In this work, we propose to use the TTS loss instead of the TTE loss, to realize 
\begin{equation}
\mathcal{L}_{\mathrm{ASR}\rightarrow\mathrm{TTS}}= E_{p_{\mathrm{ASR}}(\mathbf{Y}\mid \mathbf{X})}\{\mathcal{L}_{\mathrm{TTS}}\}\label{eq:7}.
\end{equation}
Aside from being computationally more intensive, this requires solving the problem of passing speaker characteristics through the text bottleneck, see Figure~\ref{fig:overview}(a). In order to do so, inspired by \cite{tjandra2017listening}, we augment the TTS model with speaker vectors obtained from a x-vector network~\cite{snyder2018x}. For example, for the MSE component of $\mathcal{L}_{\mathrm{TTS}}$ we have
\begin{equation}
\mathcal{L}_{\mathrm{MSE}}= -\log p_{\mathrm{TTS}}(\mathbf{X}^*\mid\mathbf{Y}, \mathbf{f}(\mathbf{X})) \label{eq:8}
\end{equation}
where $\mathbf{f}$ implements the x-vector. Note that x-vectors are designed to retain speaker characteristics but not general structure of the speech signal. In that sense, the model can not learn to copy directly $\mathbf{X}$ from input to output.

Similarly to \cite{hori2018cycle} we use policy-gradient to back-propagate through the expectation in the loss and update the ASR parameters. This yields the following gradient update formula
\begin{multline}
\nabla_\theta\mathcal{L}_{\mathrm{ASR\to TTS}} \approx \\ \sum_{\mathbf{Y}^n \sim p_{\mathrm{ASR}}} \frac{1}{N}R(\mathbf{Y}^n, \mathbf{X}) \nabla_\theta \log p_{\mathrm{ASR}}(\mathbf{Y}^n\mid\mathbf{X})
\label{REINFORCE}
\end{multline}
for $N$ samples. The reward $R$ is obtained by subtracting a bias term from the TTS loss to reduce variance $R(\mathbf{Y}^n, \mathbf{X}) = \mathcal{L}_{\mathrm{TTS}} - B(\mathbf{X})$. The bias term is calculated as the mean value of $\mathbf{X}$ for each sample. In practical terms, using policy gradient amounts to sampling multiple sentences from the ASR distribution and backpropagating each of them as if they were the ground truth, but weighted by the reconstruction loss. 

\subsection{Unsupervised training with unpaired text data}

As with speech, unsupervised training of ASR using only text and cycle-consistency is possible by using a TTS$\rightarrow$ASR chain, see Figure~\ref{fig:overview}(b). Since our target in this work is to increase ASR performance, the resulting computation graph is simpler than in the ASR$\rightarrow$TTS case. There is no need to backpropagate into the TTS module and thus the chain can be realized by forming a non-differentiable TTS$\rightarrow$ASR pipeline as
\begin{equation}
\mathcal{L}_{\mathrm{TTS}\rightarrow\mathrm{ASR}}= -\mbox{log}p_{\mathrm{ASR}}(\mathbf{Y}^{*}\mid \hat{\mathbf{X}}).\label{eq:11}
\end{equation}
The point estimate is obtained by providing the TTS system with the input text $\mathbf{Y}$ along with randomly chosen x-vectors to generate
\begin{equation}
\hat{\mathbf{X}} = \arg\max_{\mathbf{X}}\{p_{\mathrm{TTS}}(\mathbf{X}\mid \mathbf{Y})\}.
\end{equation}
Thus the pipeline can act as an autoencoder and allows training of unpaired text only data. Using a non-differentiable pipeline where TTS generates synthetic speech sentences for which we only have a transcription is akin to back-translation in machine translation \cite{sennrich2015improving}. The work in \cite{hayashi2018back} proposes a similar idea utilizing TTE instead of TTS. The TTS$\rightarrow$ASR pipeline is also used in \cite{tjandra2017listening}.

\subsection{Unsupervised training with both unpaired speech and text data}
In the case in which both paired speech and text are available, both ASR$\rightarrow$TTS and TTS$\rightarrow$ASR can be trained jointly, see figure~\ref{fig:overview}. The final loss $\mathcal{L}_{\mathrm{both}}$ is a linear interpolation of the loss functions defined in equations~\eqref{REINFORCE} and~\eqref{eq:7}. 

\begin{equation}
\mathcal{L}_{\mathrm{both}} = \alpha * \mathcal{L}_{\mathrm{ASR}\rightarrow\mathrm{TTS}} + (1 - \alpha) * \mathcal{L}_{\mathrm{TTS}\rightarrow\mathrm{ASR}}
\end{equation}

\noindent where $\alpha$ is a hyper-parameter set by default to $\alpha=0.5$.
\section{Experimental Setup}\label{sec:3}
An extensive experimental setup leveraging the Librispeech~\cite{panayotov2015librispeech} and Wall Street Journal (WSJ)~\cite{paul1992design} corpora was used in this work. For each corpus, a sub-set of the data was used to simulate unpaired speech and text conditions.

Training utilized 83-dimensional filter-bank with pitch features as input. 
The encoder-decoder network utilized location aware attention~\cite{bahdanau2016end}. For WSJ and Librispeech experiments, the encoder comprises 8 bi-directional LSTM layers~\cite{schuster1997bidirectional,hochreiter1997long} each with 320 units and the decoder comprises 1 (uni-directional) LSTM layer with 320 units. The CE objective is optimized using AdaDelta \cite{zeilerAdaDelta} with an initial learning rate set to $1.0$. The training batch size is 30 and the number of training epochs is 20. The learning rate decay is based on the validation performance computed using the character error rate (min. edit distance). ESPnet \cite{watanabe2018espnet} is used to implement and execute all our experiments.
For TTS, Tacotron2~\cite{shen2018natural} model is used and is configured as described in our previous work~\cite{hori2018cycle} differing only with the output targets. Here, we use 83 dimensional log-Mel filterbank features as targets and x-vectors~\cite{snyder2018x} are fed as auxiliary information to provide speaker characteristics of each utterance.
Detailed experimental setup can be obtained from our github codebase in ESPnet. 
Unsupervised training was performed after conventional supervised training of each model.
Cycle-consistency utilized five samples in Eq.~\ref{REINFORCE}. A small amount of paired data was also used to regularize the model during the unsupervised stage. For WSJ, 81 hours were split into 2, 5, 10 and 14 hours of paired data and the remaining 67 hours was used as unpaired data. Table~\ref{tab:pretrain} shows the effect of the amount of parallel data on WSJ and Librispeech using conventional supervised training regime. The eval92 test set was kept for evaluations. In Librispeech, the 100 hours is used as paired data and the 360 hours set as unpaired data as in~\cite{hori2018cycle}. 
\begin{table}[H]
\caption{Baseline performance of paired data (semi-supervised) across WSJ-SI84 and Librispeech by using pre-trained and randomly initialized ASR model on eval-92 test for WSJ and test-clean for Librispeech}\label{tab:pretrain}
\centering{}{\footnotesize{}}%
\begin{tabular}{llcc}
\hline 
\multicolumn{4}{c}{{\footnotesize{}Paired data Corpus info  }}\tabularnewline
{\footnotesize{}Name} & {\footnotesize{}\# Hours} & {\footnotesize{}\% CER} & {\footnotesize{}\% WER}\tabularnewline
\hline 
\hline 
\multirow{4}{*}{{\footnotesize{}WSJ}} & {\footnotesize{}2} & {\footnotesize{}27.7} & {\footnotesize{}68.2}\tabularnewline
 & {\footnotesize{}5} & {\footnotesize{}13.2} & {\footnotesize{}41.5}\tabularnewline
 & {\footnotesize{}10} & {\footnotesize{}10.8} & {\footnotesize{}33.7}\tabularnewline
 & {\footnotesize{}14} & {\footnotesize{}10.2} & {\footnotesize{}31.5}\tabularnewline
{\footnotesize{}Librispeech} & {\footnotesize{}100} & {\footnotesize{}8.9} & {\footnotesize{}21.0}\tabularnewline
\hline 
\end{tabular}{\footnotesize \par}
\end{table}

\section{Analysis and Results}\label{sec:4}
The initial experiments compared the performance of the TTS loss introduced in this work, to the TTE loss from previous work~\cite{hori2018cycle}. 
The analysis involved Librispeech setup using 100 hours of paired data with 180 and 360 hours of unpaired data for a fair comparison. The results shows that TTS approach improves \%WER over TTE by a 2.5\% relative (20.7\% to 20.1\%) on 360 hours of unpaired data and 2.9\% relative (19.9\% to 19.4\%) on 180 hours set. 
\subsection{Impact of amount of paired data (semi-supervision)}
In this section, the model is tested with varying amounts of unpaired data such as 14 hours and 67 hours along with certain amount of semi-supervision such as 2, 5, 10 and 14 hours of data.  
Table~\ref{tab:wsj}, shows the \%WER obtained by using varying amounts of unpaired speech and text data and different amounts of paired data.  The experiments are done starting from 2 hours of parallel data and 14 hours of unpaired data. With only 2 hours of parallel data, the model \%WER performance improves from 68.2 in Table~\ref{tab:pretrain} to 49.8 by only aiding 14 hours of unpaired speech and improved to 51.9 with 67 hours of unpaired speech.

Overfitting is observed in the case of unpaired speech, while in case of text only data, the model improved consistently from 63 (pair 14 hours) to 39.6 (pair 67 hours). This showcases the importance of text only data under very low-resource scenario. Interestingly,  including both unpaired speech and text of 14 hours, the \%WER improved to 43.7 and with 67 hours of data the model obtained 41.4 \%WER. The pattern emerging here is that, under very low-resource scenario, the model benefits from large amounts of text. But, adding more speech only data leads to slight degradation in performance. This pattern is not observed with 5, 10 and 14 hours of parallel data conditions as increasing the amount of speech data from 14 hours to 67 hours improved by $\sim$1\% absolute WER. With 14 hours of supervision and 67 hours of unpaired speech, text and both, the \%WER improves to 28.0, 27.1 and 26.2 respectively.
\begin{table}[H]
\caption{\%WER performance analysis on varying the amount of paired data (semi-supervision) and unpaired data on eval-92 test set using WSJ. The Type refers to type unpaired data used and it can be either "speech/text/both"}\label{tab:wsj}
\centering{}%
\begin{tabular}{lccccc}
\hline 
\multicolumn{2}{c}{\footnotesize{}Unpaired data} & \multicolumn{4}{c}{\footnotesize{}Paired data (\#hrs)}\tabularnewline
\cline{3-6} 
{\footnotesize{}\#hrs} & {\footnotesize{}Type} & {\footnotesize{}2} & {\footnotesize{}5} & {\footnotesize{}10} & {\footnotesize{}14}\tabularnewline
\hline 
\hline 
{\footnotesize{}14}  & {\footnotesize{}Speech} & {\footnotesize{}49.8} & {\footnotesize{}39.9} & {\footnotesize{}29.8} & {\footnotesize{}-}\tabularnewline

{\footnotesize{}14}  & {\footnotesize{}Text} & {\footnotesize{}63.0} & {\footnotesize{}43.6} & {\footnotesize{}34.6} & {\footnotesize{}-}\tabularnewline

{\footnotesize{}14}  & {\footnotesize{}Both} & {\footnotesize{}43.7} & {\footnotesize{}35.5} & {\footnotesize{}28.3} & -\tabularnewline

{\footnotesize{}67}  & {\footnotesize{}Speech} & {\footnotesize{}51.9} & {\footnotesize{}38.8} & {\footnotesize{}28.4} & {\footnotesize{}28.0}\tabularnewline

{\footnotesize{}67}  & {\footnotesize{}Text} & {\footnotesize{}39.6} & {\footnotesize{}36.8} & {\footnotesize{}29.6} & {\footnotesize{}27.1}\tabularnewline

{\footnotesize{}67}  & {\footnotesize{}Both} & {\footnotesize{}41.4} & {\footnotesize{}34.2} & {\footnotesize{}27.7} & {\footnotesize{}26.2}\tabularnewline
\hline 
\end{tabular}
\end{table}

\begin{table}[H]
\caption{Unsupervised ASR performance across best results in literature. The Type refers to type unpaired data used and it can be either "speech/text/both"}\label{tab:lit}
\centering{}%
\resizebox{\columnwidth}{!}{
\begin{tabular}{lcccc}
\hline 
\multicolumn{5}{c}{{\footnotesize{}WSJ-SI84 (parallel) + WSJ-SI284 (unpaired)}}\tabularnewline
\hline 
{\footnotesize{}Model} & {\footnotesize{}Type} & {\footnotesize{}RNNLM} & {\footnotesize{}\%CER} & {\footnotesize{}\%WER}\tabularnewline
\hline 
\hline 
{\footnotesize{}Speech chain~\cite{tjandra2017listening}} & {\footnotesize{}Both} & - & {\footnotesize{}9.9} & -\tabularnewline
{\footnotesize{}Adversarial~\cite{drexler2018combining}}& {\footnotesize{}Both} & {\footnotesize{}yes} & - & {\footnotesize{}24.9}\tabularnewline
 {\footnotesize{}this work} & {\footnotesize{}Both} & - & {\footnotesize{}9.1} & {\footnotesize{}26.1}\tabularnewline
 {\footnotesize{}this work} & {\footnotesize{}Both} & {\footnotesize{}yes} & {\footnotesize{}\textbf{7.8}} & {\footnotesize{}\textbf{20.3}}\tabularnewline
 {\footnotesize{}oracle} & - & - & {\footnotesize{}5.5} & {\footnotesize{}16.4}\tabularnewline
 {\footnotesize{}oracle~\cite{baskar2018promising}} & - & yes & {\footnotesize{}2.0} & {\footnotesize{}4.8}\tabularnewline
\hline
\multicolumn{5}{c}{{\footnotesize{}Librispeech 100 h (parallel) + 360 h (unpaired)}}\tabularnewline
\hline 
\hline 
{\footnotesize{}Backtranslation-TTE~\cite{hayashi2018back}}& {\footnotesize{}Text}  &- & {\footnotesize{}10.0} & {\footnotesize{}22.0}\tabularnewline
{\footnotesize{}this work} & {\footnotesize{}Text} & {\footnotesize{}-} & {\footnotesize{}8.0} & {\footnotesize{}17.9}\tabularnewline
{\footnotesize{}Crictizing-LM~\cite{liuLM}}& {\footnotesize{}Text} & {\footnotesize{}yes} & {\footnotesize{}9.1} & {\footnotesize{}17.3}\tabularnewline
{\footnotesize{}this work} & {\footnotesize{}Text} & {\footnotesize{}yes} & {\footnotesize{}8.0} & {\footnotesize{}17.0}\tabularnewline
{\footnotesize{}Cycle-TTE~\cite{hori2018cycle}} & {\footnotesize{}Speech} & {\footnotesize{}yes} & {\footnotesize{}9.9} & {\footnotesize{}19.5}\tabularnewline
{\footnotesize{}this work} & {\footnotesize{}Speech} & {\footnotesize{}yes} & {\footnotesize{}7.8} & {\footnotesize{}16.8}\tabularnewline
{\footnotesize{}Adversarial-AE~\cite{karita2018sequence}} & {\footnotesize{}Both} & {\footnotesize{}yes} & {\footnotesize{}8.4} & {\footnotesize{}18.0}\tabularnewline
{\footnotesize{}this work} & {\footnotesize{}Both} & - & {\footnotesize{}7.6} & {\footnotesize{}17.5}\tabularnewline
{\footnotesize{}this work} & {\footnotesize{}Both} & {\footnotesize{}yes} & {\footnotesize{}\textbf{7.6}} & {\footnotesize{}\textbf{16.6}}\tabularnewline
{\footnotesize{}oracle~\cite{hori2018cycle}} & - & - & {\footnotesize{}4.6} & {\footnotesize{}11.8}\tabularnewline
\hline
\end{tabular}}
\end{table}
\subsection{Model comparison with literature}
Table~\ref{tab:lit}, shows the results of using both unpaired speech and text data from WSJ-SI284 and Librispeech 360 hours across literature. In WSJ corpus, the model achieves a \%WER of 20.3 with RNNLM and 26.1 without RNNLM. This leaves a relative difference of 37.1\% compared to oracle performance of 16.4\%. The oracle result is our baseline performance using WSJ-SI284. Note that the performance on WSJ-SI284 is inferior to our previously reported baseline~\cite{baskar2018promising}. This is due to a difference in architecture necessary to fit ASR and TTS models into the GPU.
In case of Librispeech, the table~\ref{tab:lit} shows that training with unpaired audio and text data can achieve a \%WER of 17.5 leading to 32.5\% relative difference with the oracle performance. Table~\ref{tab:lit} also shows that our approach only using unpaired text gains 18.0\% relative improvement over backtranslation-TTE~\cite{hayashi2018back} approach. Complementary gains of absolute 0.9\% were observed by integrating RNNLM with this approach and the result is compared with crictizing-LM~\cite{liuLM}. The effectiveness of our approach using unpaired speech only data is shown in table~\ref{tab:lit} by obtaining 16.8\% WER. Jointly training unpaired speech and text provided modest gains with a \%WER improvement from 16.8 to 16.6. From these results, one can infer that training with unpaired speech only data has major benefits over text only data in large corpus such as Librispeech.

\section{Related Work}
As indicated in the previous sections, the work here proposed improvements over recent related approaches and integrated some of them into a single loss. Back-translation style TTE~\cite{hayashi2018back} synthesized encoder can be related to the TTS$\rightarrow$ASR loss here used, as both utilize a pipeline that generates synthetic train data. The latter has the advantage of utilizing directly speech and has higher performance overall. The speech-chain framework~\cite{tjandra2017listening} was the first work to integrate ASR and TTS to train using unpaired speech and text. It also utilizes a TTS$\rightarrow$ASR pipeline loss but this is not learnt jointly with the end-to-end differentiable ASR$\rightarrow$TTS loss. One limitation of the speech-chain is the fact that is not end-to-end differentiable, having to rely in the alternate training of a TTS$\rightarrow$ASR and ASR$\rightarrow$TTS loss to update both models. This limitation was addressed in \cite{hori2018cycle} proposing an end-to-end differentiable loss for speech only and based on TTE. The work presented here improves upon this by extending TTE to TTS with x-vectors and combining this with a TTS$\rightarrow$ASR point-estimate loss. Finally~\cite{tjandra2018end} is similar to \cite{hori2018cycle} but applies straight-through estimators to construct end-to-end differentiable ASR$\rightarrow$TTS losses based on $\mathrm{argmax}$ and a expected loss similar to Eq.~\ref{eq:7}.

\section{Conclusions}
This paper presented a new approach to exploit the information in unpaired speech and/or unpaired text to improve the performance of seq2seq ASR systems. We show that under low-resource conditions such as WSJ corpus the performance improvements are relatively higher compared to corpus with sufficient amount of data such as Librispeech. We show as well that integrating unpaired speech and text, both as a pipeline loss and through shallow integration with a RNN language model, provides additional gains and competitive results. Future work will be focused on cycle-consistency approaches where ASR and TTS do not have matching conditions. Preliminary experiments show that this is a limitation of current systems.

\section{Acknowledgements}
This work was supported by Czech Ministry of Education, Youth and Sports from the National Program of Sustainability (NPU II) project "IT4Innovations excellence in science - LQ1602" and by the Office of the Director of National Intelligence (ODNI), Intelligence Advanced Research Projects Activity (IARPA) MATERIAL program, via Air Force Research Laboratory (AFRL) contract \# FA8650-17-C-9118 and U.S. DARPA LORELEI contract No. HR0011-15-C-0115. The views and conclusions contained herein are those of the authors and should not be interpreted as official policies, either expressed or implied, of ODNI, IARPA, AFRL or the U.S. Government. The work was also supported by Facebook (Research Award on Speech and Audio Technology for Voice Interaction and Video Understanding.

\bibliographystyle{IEEEtran}
\bibliography{ref_new}{}


\end{document}